\documentclass[final,5p,times,twocolumn]{elsarticle}

 \usepackage{graphicx}

\usepackage{amssymb}

\journal{NIM A  RICAP-2013}

\begin{document}

\begin{frontmatter}



\title{Recent results from the ANTARES neutrino telescope}


\author[VVE]{V\'eronique Van Elewyck}

\address[VVE]{APC, Universit\'e Paris Diderot, CNRS/IN2P3, CEA/Irfu, Obs. de Paris, Sorbonne Paris Cit\'e, France}

\author{ \\ on behalf of \\ the ANTARES Collaboration}

\begin{abstract}
The ANTARES neutrino telescope is currently the largest operating water Cherenkov detector and the largest neutrino detector in the Northern Hemisphere.  
Its main scientific target is the detection of high-energy (TeV and beyond) neutrinos from cosmic accelerators, as predicted by hadronic interaction models, and the measurement of the diffuse neutrino flux. Its location allows for surveying a large part of the Galactic Plane, including the Galactic Centre.

In addition to the standalone searches for point-like and diffuse high-energy neutrino signals, ANTARES has developed a range of multi-messenger strategies to exploit the close connection between neutrinos and other cosmic messengers such as gamma-rays, charged cosmic rays and gravitational waves. This contribution provides an overview of the recently conducted analyses, including a search for neutrinos from the Fermi bubbles region, searches for optical counterparts with the TAToO program, and searches for neutrinos in correlation with gamma-ray bursts, blazars, and microquasars. 
Further topics of investigation, covering e.g. the search for neutrinos from dark matter annihilation, searches for exotic particles and the measurement of neutrino oscillations, are also reviewed.
\end{abstract}


\end{frontmatter}


\section{Introduction}

Neutrinos have long been proposed as an alternative to cosmic rays and photons to explore the high-energy sky, as 
they can emerge from dense media and travel across cosmological distances without being deflected by magnetic fields nor absorbed by ambient matter and radiation. High-energy ($>$TeV) neutrinos are expected to be emitted in a wide range of astrophysical objects. Along with gamma-rays, they originate from the decay of $\pi$'s and $K$'s produced in the interactions of accelerated protons and nuclei with matter and radiation in the vicinity of the source. Candidate neutrino emitters therefore include most known or putative cosmic ray accelerators, ranging from galactic sources such as supernovae remnants or microquasars to the most powerful extragalactic emitters such as Active Galactic Nuclei (AGNs) and Gamma-Ray Bursts (GRBs)~(see~\cite{Becker_review2008} for a review on the subject). Apart from tracing the existence of hadronic processes inside astrophysical sources, the observation of cosmic neutrinos could also reveal new types of sources, yet unobserved in photons or cosmic rays. Due to oscillations, the expected flavor ratio at Earth for cosmic neutrinos is $\nu_e : \nu_\mu : \nu_\tau = 1 : 1 : 1$. 

Neutrino astronomy is challenged both by the weakness of neutrino interactions and by the feebleness of the expected fluxes of cosmic neutrinos.  Neutrino telescopes therefore require the instrumentation of large volumes of transparent medium (water or ice) with arrays of photosensors, in order to detect the Cherenkov light emitted by the charged leptons produced when a neutrino interacts with material in or around the detector.  Muons are the most straightforward detection channel, but showers induced by electron- and tau-neutrino can also be detected. The timing, position and amplitude of the light pulses (or hits) recorded by the photosensors allow the  reconstruction of the muon trajectory, providing the arrival direction of the parent neutrino and an estimation of its energy. Such detectors are installed at great depths and optimized to detect up-going muons produced by neutrinos which have traversed the Earth, in order to limit the background from down-going atmospheric muons. As a consequence, one single neutrino telescope can efficiently monitor only half of the sky in the TeV -- PeV range. The ANTARES detector, located in the Mediterranean Sea, is well suited for the observation of Galactic sources; its layout and performances are described in Sec.~\ref{sec:antares}.

Atmospheric neutrinos produced in cosmic-ray-induced air showers can also traverse the Earth and interact close to the detector, providing an irreducible source of background with energy spectrum $\propto E_\nu^{-3.7}$. Diffuse fluxes of neutrinos from astrophysical origin, which are expected to be harder (typically $\propto E_\nu^{-\alpha}$ with $\alpha = $ 1\,--\,2), can then be identified as an excess of events above a certain energy. 
Another approach consists in looking for localized excesses of events, either blindly over the full sky, or in the direction of a-priori selected candidate source locations that correspond e.g. to known gamma-ray emitters. Such searches are well suited to look for steady, point-like sources, in particular in the Galaxy. Finally, multimessenger searches develop specific strategies to look for neutrinos with timing and/or directional correlations with other cosmic messengers such as photons, charged cosmic rays or gravitational waves. The outcome of such searches performed with the ANTARES detector are described in Secs.~\ref{sec:diffuse}~(diffuse fluxes), \ref{sec:point}~(neutrino sources) and ~\ref{sec:mm}~(multimessenger searches).

Beyond its astrophysics goals, the science reach of a neutrino telescope such as ANTARES also includes an extended particle physics program, ranging from the study of neutrino oscillations and the indirect search for dark matter (in the GeV -- TeV energy range) to the search for exotic particles (at the upper end of the energy spectrum). These topics are discussed in Sec.~\ref{sec:beyond}.  Recent reviews on the techniques and challenges of neutrino astronomy can be found e.g. in ~\cite{Katz_review2012,Ribordy_review2012}.

\section{The ANTARES detector}
\label{sec:antares}

 \begin{figure}[!t]
\includegraphics[width=0.48\textwidth]{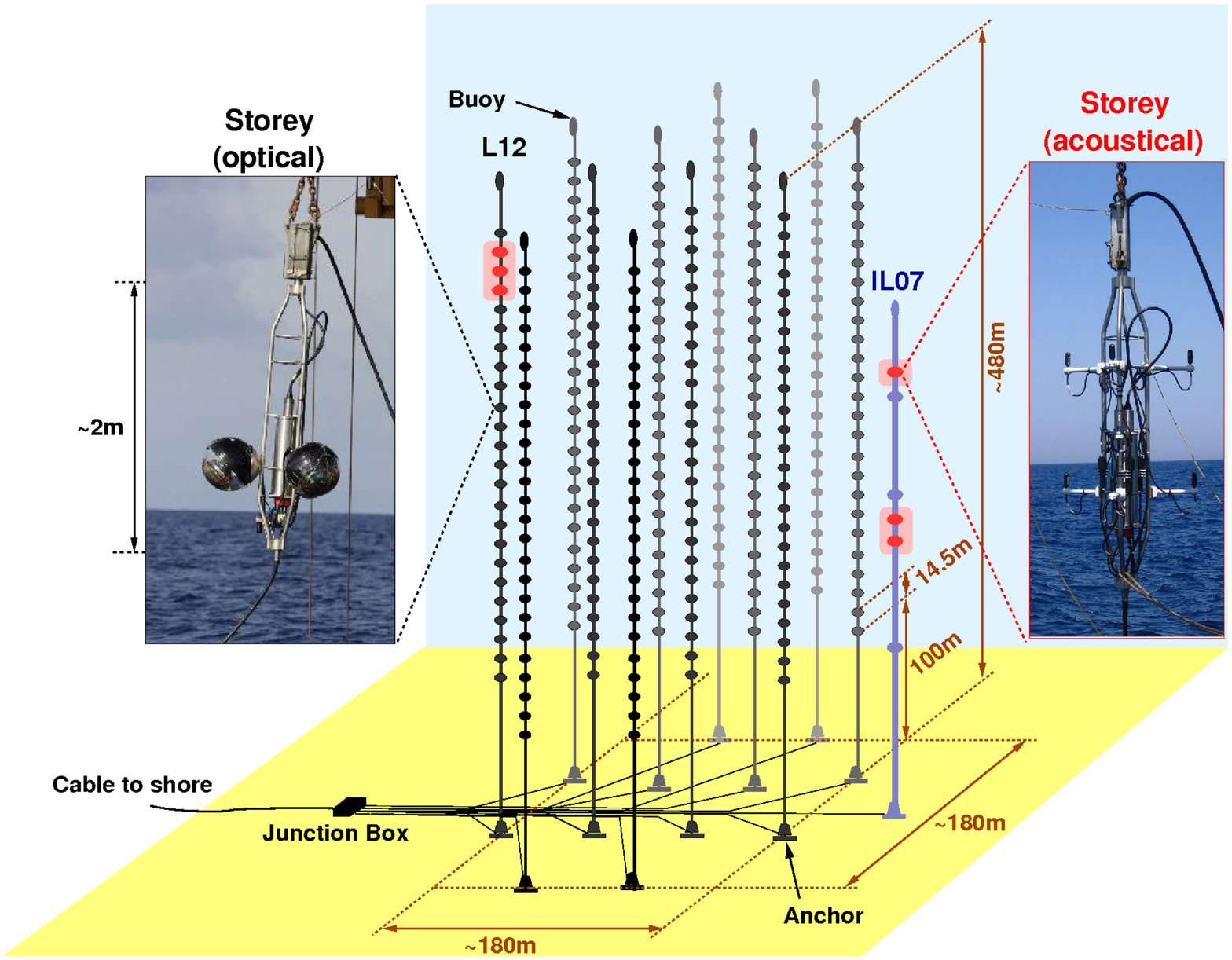}
\caption{The ANTARES detector configuration, showing the 12 lines hosting standard storeys with optical modules (in black), as well as the instrumented line (IL07). The red storeys on L12 and IL07 correspond to the AMADEUS acoustic detection modules.}
\label{fig:antares}
\end{figure}

The ANTARES (Astronomy with a Neutrino Telescope and Abyss environmental Research) detector is the first undersea neutrino telescope and the only one currently operating~\cite{ANTARES_detector}. Its deployment at a depth of 2475m in the Mediterranean Sea, about 40 km off the French coast near Toulon (Var), was completed in May 2008. It consists of an array of 12 lines anchored to the sea bed and connected to a junction box which distributes the electrical power and transmits the data to shore through an electro-optical cable (see Fig.~\ref{fig:antares}). The detector lines are 450m high, with an inter-line spacing ranging from 60m to 75m. Each line supports 25 storeys made of triplets of 10-inch photomultipliers (PMTs) enclosed in 17-inch pressure-proof glass spheres. The detector comprises a total of 885 optical modules (OMs), corresponding to an instrumented volume of about 0,02 km$^3$. Acoustic devices and inclinometers regularly spread along the lines allow to accurately monitor the position and orientation of the OMs~\cite{positioning}, and time calibration is performed by means of an {\it in situ} array of laser and LED beacons~\cite{timecalib}. 

The PMTs are orientated at $45^\circ$ downwards in order to maximize the sensitivity to Cherenkov light from upcoming muons. The time and charge of all PMT signals (or ``hits'') that pass a predefined threshold (typically 0.3 photoelectron) are digitized and sent to shore, where they are processed on the fly by a PC farm. The data flow to the shore amounts to about 1 to 10 Gb/s; it is dominated by random hits ($\sim$ 60 -- 150 kHz per PMT) due to the background light  produced by the radioactive decay of $^{40}$K and by bioluminescence. Online filtering algorithms run on the PCs to select the physics events among the raw data and write them to disk to be analysed offline. Such events are then reconstructed using the hits time and position information, typically by means of a modified maximum likelihood method~\cite{reco}. The median angular resolution achieved for muon tracks is  $\lesssim 0.5^\circ$, allowing good performance in the searches for neutrino point sources, as reported in Sec.~\ref{sec:point}.  

The detector records a baseline rate of $\sim$10 Hz of muons. Those associated with upgoing (atmospheric) neutrinos can be identified by imposing both quality cuts on the track fit parameters, and a directional cut on the reconstructed zenith angle ($\theta>90^\circ$). Within these conditions, the detector instantaneous sky coverage is $2 \pi$ sr, and its integrated visibility covers about $\frac{3}{4}$ of the sky. Its location in the Northern Hemisphere allows for surveying a large part of the Galactic Plane, including the Galactic Centre, thus complementing the sky coverage of the IceCube detector installed at the South Pole~\cite{icecube}. 

The main junction box of ANTARES also supports an additional instrumented line which provides measurements of environmental parameters such as sea current and temperature, and hosts part of the AMADEUS prototype array for ultra-high energy neutrino acoustic detection~\cite{AMADEUS_NIM2011}. Since April 2013, this line hosts a prototype OM,  housing  31 3-inch PMTs, designed for the next-generation deep-sea neutrino telescope, KM3NeT~\cite{km3net}. A secondary junction box dedicated to Earth and Marine Science projects was also connected to the main junction box and is operating since December 2010.

\section {Searches for diffuse neutrino fluxes}
\label{sec:diffuse}
\subsection{Measurement of the atmospheric $\nu_\mu$ spectrum}
\label{subsec:atmo}
 The bulk of atmospheric neutrinos detected by ANTARES represent an irreducible background in the search for cosmic neutrinos. A precise measurement of this atmospheric neutrino flux is nevertheless of intrinsic interest both as a test of the good behaviour of the detector, and as a valuable source of information on the physics of cosmic-ray-induced atmospheric showers. Up to about 100 TeV, the dominant neutrino production channels are $\pi^\pm$ and $K^\pm$ decays, yielding an energy spectrum $\sim E^{-3,7}$. At higher energies, an additional component originating from the prompt decay of charmed and bottomed particles takes over, with a spectrum which is expected to follow more closely that of the primary cosmic rays ($\sim E^{-2,7}$). Its precise features are however sensitive to the hadronic interaction models, which are still poorly constrained at those high energies.

A measurement of the atmospheric $\nu_\mu$ spectrum has been performed using the ANTARES data collected between December 2007 and December 2011 (hereafter dubbed as the 2008-2011 data sample), for a total equivalent live time of 855 days. The cuts used to select a pure sample of atmospheric neutrinos  were optimized using a run-by-run Monte Carlo simulation which reproduces the detail of the acquisition conditions of the detector. 10\% of the data were used as a burn sample to check the agreement between the observed and the expected distributions. 

The determination of the neutrino energy is a crucial -- and challenging -- aspect of atmospheric neutrino spectrum studies. In the present analysis, two energy estimators were used. The first one is based on a maximum likelihood approach attempting to maximize the agreement between the observed and expected amount of detected light; the second one relies on the muon energy loss along its trajectory. Nevertheless, the limited resolution of these estimators make it impossible to reconstruct the neutrino energy spectrum on an even-by-event basis. Instead, an unfolding procedure based on Monte Carlo simulations is used; more detail on the analysis can be found in~\cite{antares_nuatm}.

\begin{figure}[!t]
\includegraphics[width=0.48\textwidth]{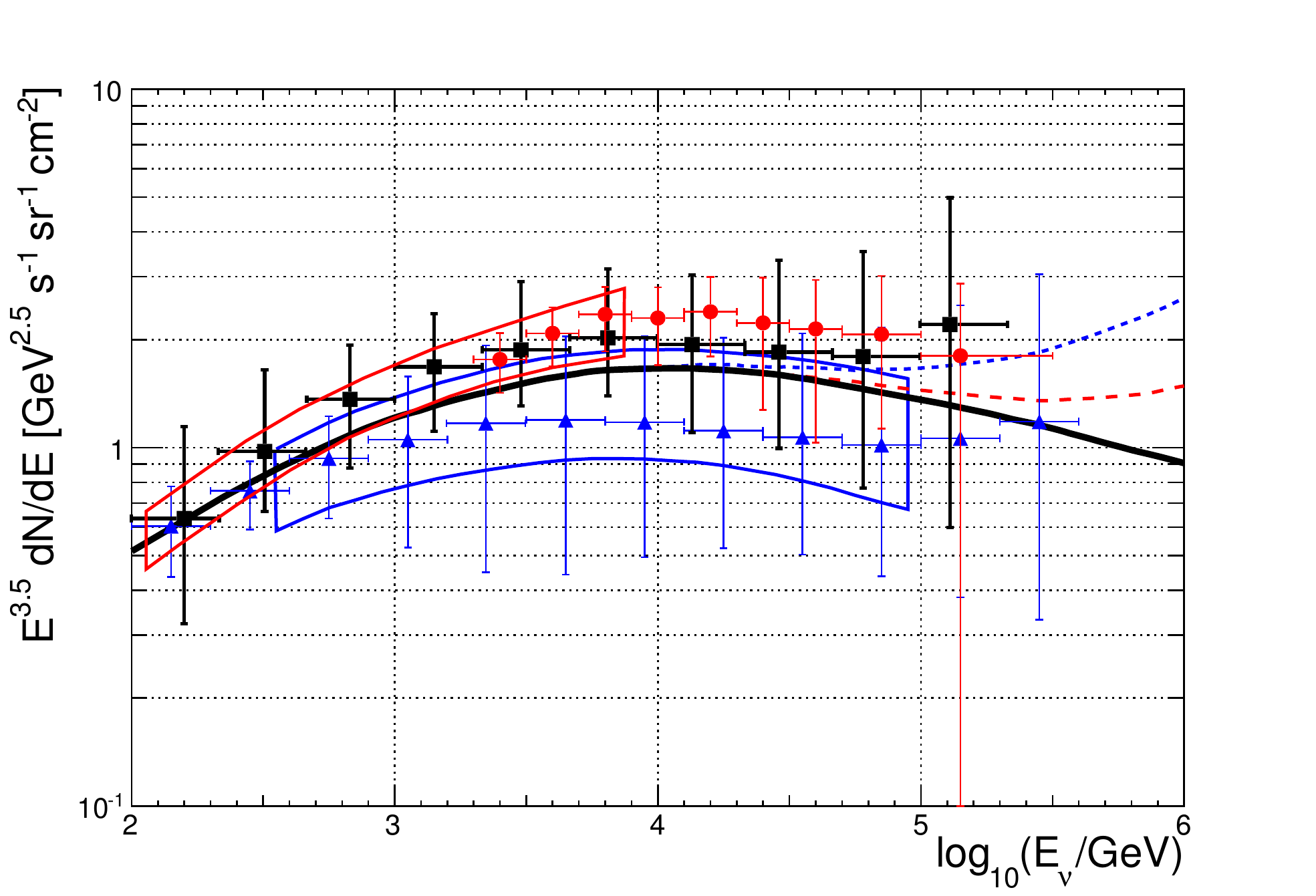}
\caption{Zenith-averaged atmospheric neutrino spectrum multiplied by $E_\nu^{3.5}$. The ANTARES measurement (black squares) is shown together with the results from AMANDA-II (red circles)~\cite{amanda_nu} and and IceCube 40 (blue triangles)~\cite{icecube40_nu}. The black, solid line represents the conventional flux prediction from the Bartol group~\cite{bartol} ; the red and blue dashed lines include two prompt neutrino production models from~\cite{prompt1} and~\cite{prompt2}, respectively. The
red region corresponds to the older measurement from AMANDA-II~\cite{amanda_nu2}, and the blue one the IC40 update from~\cite{icecube40_nu2}. 
}
\label{fig:nuatm}
\end{figure}

The unfolded neutrino energy spectrum, averaged over zenith angles $\theta > 90^\circ$, is shown in Fig.~\ref{fig:nuatm} as a function of the energy in the range 100 GeV -- 200 TeV. The decrease above $E_\nu \sim 100$ TeV is expected from the change of the spectral index of the primary cosmic ray flux above the knee region
($E_{knee} \simeq 3\times 10^{15}$ eV). It is about 25\% higher than the prediction from the Bartol group~\cite{bartol}, though still well within uncertainties, and intermediate between the spectra measured by the Antarctic neutrino telescopes AMANDA-II and IceCube-40. This good overall agreement provides confidence in the understanding of the absolute energy scale of the detector.

\subsection{Search for a diffuse neutrino flux of astrophysical origin}

\begin{figure}[!t]
\hspace*{-0.5cm}
\includegraphics[width=0.50\textwidth]{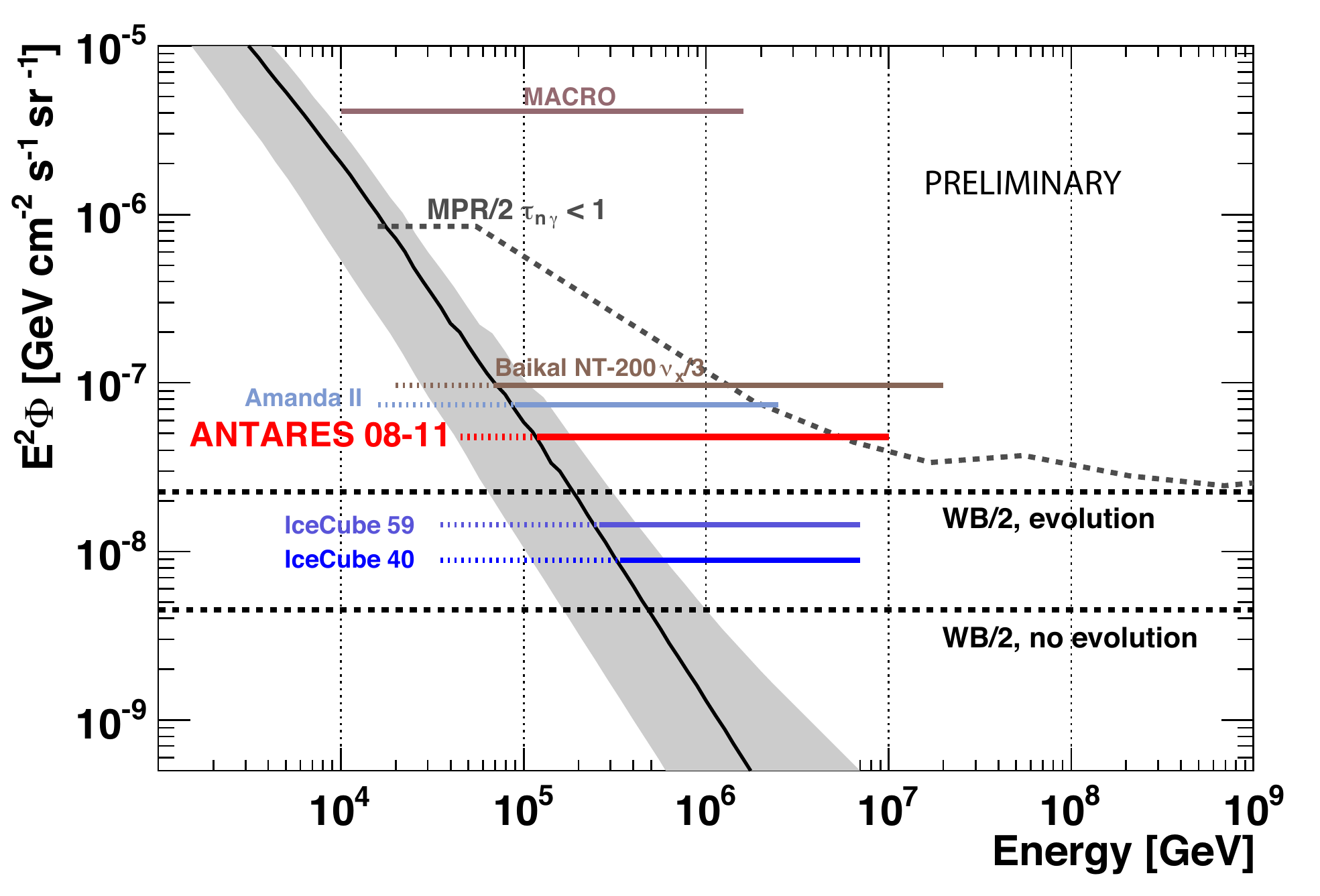}
\caption{Updated ANTARES upper limit at 90\% C.L. for an $E^{-2}$ spectrum, compared to 90\% C.L. limits from other experiments, including IceCube in its 40~\cite{icecube40_nu2} and 59-string~\cite{ic59_diffuse} configurations. Also shown are the Waxman-Bahcall benchmark flux predictions~\cite{wb}.}
\label{fig:diffuse}
\end{figure}

Once the atmospheric neutrino flux has been measured, one can look for an excess of events, that would reflect the existence of a diffuse neutrino flux from the bulk of unresolved astrophysical sources. This excess would become apparent above a certain energy threshold, depending on the absolute normalization and spectral index of the cosmic neutrino flux. 

An updated analysis using the 2008-2011 data sample has been performed with the same statistical method that was used for the first (2008-2009) ANTARES limit~\cite{antares_diff1}. The main difference lies in the use of an energy estimator based on the muon energy loss along its path, $\frac{dE}{dx}$, as evaluated from the total collected charge of the hits that are selected to enter the track fitting procedure. An optimisation procedure based on the Model Rejection Factor (MRF) approach was then applied on Monte Carlo simulations (again validated on a burn sample of 10\% of the data), in order to determine the position of the energy cut yielding the best sensitivity. In view of the $\sim$25\% overall excess of the data with respect to the conventional atmospheric flux expectation (as described in Sec.~\ref{subsec:atmo}, the background was conservatively normalized to data. 

After normalization, 8 events were observed in the selected high-energy region, while 8.5 were expected from the atmospheric neutrino  background. This result yields a 90\% confidence level (C.L.) upper limit on a diffuse astrophysical component of \begin{equation}
E_\nu^2 \frac{dN}{dE_\nu} = 4.8\times 10^{-8}$ GeV cm$^{-2}$ s$^{-1}$ sr$^{-1}
\end{equation}
 in the energy range 45 TeV -- 10 PeV. This updated limit, shown in Fig.~\ref{fig:diffuse} for an $E_\nu^{-2}$ spectrum is only a factor of $\sim 4$ above the limit obtained by IceCube in its 59-string configuration~\cite{ic59_diffuse}, and it is approaching the benchmark flux predictions from~\cite{wb}. Complementarily to IceCube, it applies to the Southern sky.

\subsection{Search for neutrinos from the Fermi bubbles}
The Fermi-LAT Collaboration has recently reported the observation of an excess of gamma-rays from two extended, "bubble"-like regions above and below the Galactic Plane. The observed signal is characterized by a homogeneous intensity over the whole bubble regions and a hard ($\sim E^{-2}$) spectrum with a probable cutoff at energies above 100 GeV. The origin of these so-called Fermi bubbles (FBs) is still debated. Several proposed scenarios invoke hadronic mechanisms in which accelerated cosmic rays interact with the interstellar medium, producing pions which would in turn generate neutrinos along the observed gamma-rays~\cite{galwind,had1,had2}. 

\begin{figure}[!t]
\hspace*{-0.8cm}
\includegraphics[width=0.55\textwidth]{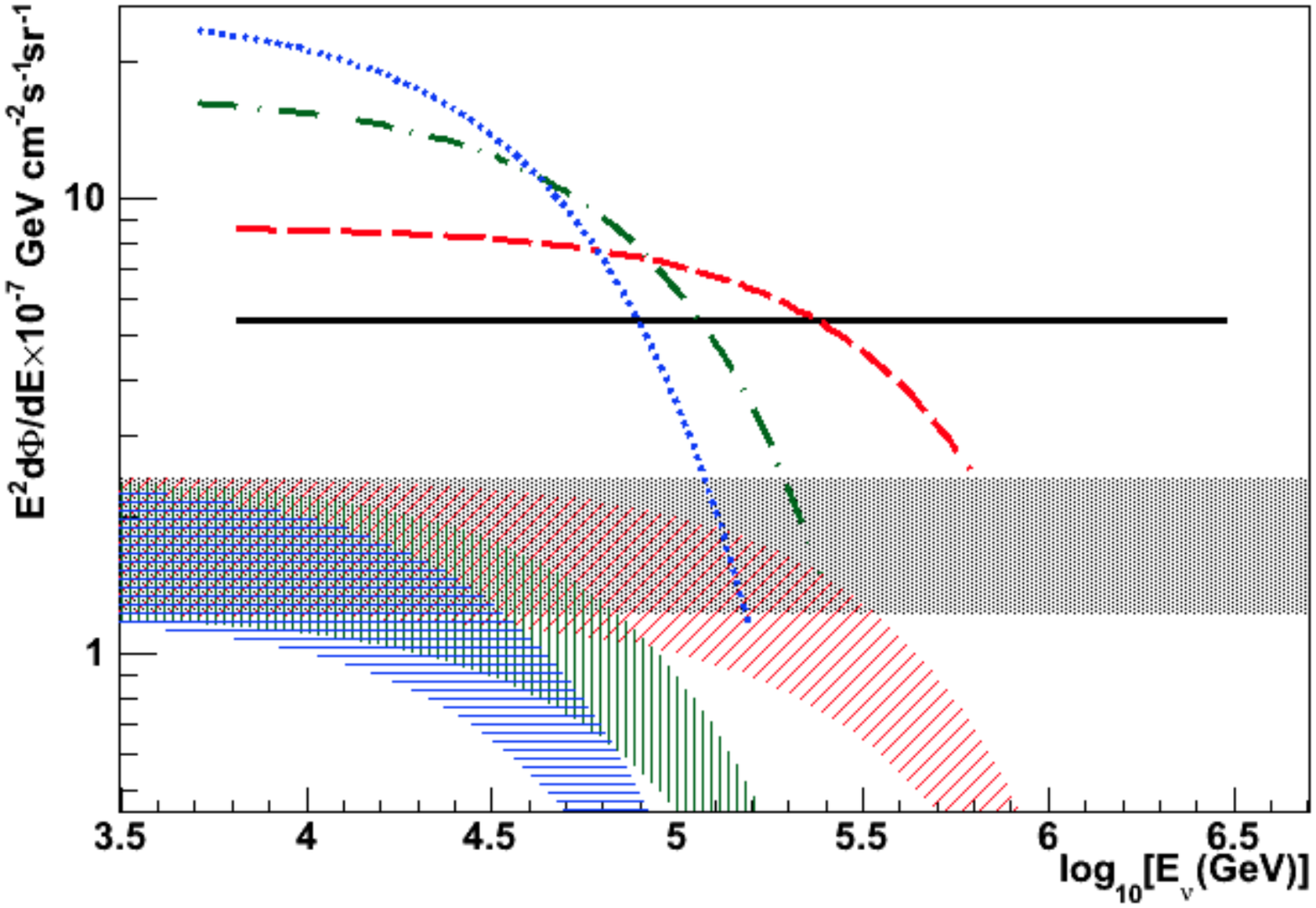}
\vspace*{-1cm}
\caption{Upper limits on the neutrino flux from the Fermi bubbles assuming different energy cutoffs at the source: no cutoff (black solid), 500 TeV (red dashed), 100 TeV (green dot-dashed), 50 TeV (blue dotted), together with the theoretical predictions for the case of a purely hadronic model (the same colours, areas  filled with dots, inclined lines, vertical lines and
horizontal lines respectively). The limits are drawn for the energy range where 90\% of the signal is expected.}
\label{fig:fermi}
\end{figure}

A dedicated search has been performed with the 2008-2011 data sample of ANTARES, taking advantage of the good sensitivity of the detector to the central region of our Galaxy; the detail of the analysis can be found in~\cite{antares_icrc_fb}. The rate of high-energy neutrino events observed in the FB region (ON zone) has been compared to that observed in equivalent (in size, shape and average detector efficiency) areas of the Galaxy excluding the FBs (OFF zones). No statistically significant excess of events above the chosen energy cut was observed in the ON zone with respect to the average expected number from the OFF zones. Corresponding upper limits on the neutrino flux from the FBs have therefore been derived for different assumptions on the energy cutoff at the source, as shown in Fig.~\ref{fig:fermi}. Some of these limits are relatively close to the expected fluxes, suggesting that at least part of the phase space for hadronic models of gamma-ray emission in the FBs could be probed with the full ANTARES dataset, or after about one year of operation of the next-generation KM3NeT neutrino telescope~\cite{km3net_fb}.

\section{Searches for sources of cosmic neutrinos}
\label{sec:point}

Such searches aim at detecting significant excesses of events from particular spots (or small regions) of the sky. They can be performed either blindly over the full sky, or in the direction of a-priori selected candidate source locations that correspond e.g. to known gamma-ray emitters. They are well-suited to look for steady, point-like sources, in particular in the Galaxy. 

These searches have first been performed on a data sample covering the period from February 2007 (when the detector was still under construction and operating with only 5 lines) until mid-November 2010, for a total equivalent live time of 813 days~\cite{antares_ps}.  The studies presented here use an updated dataset extending until end of December 2012, for a total live time of 1339 days. 
The final neutrino sample is obtained after tight cuts on the reconstruction quality parameter, on the estimated angular resolution ($<1^\circ$) and on the zenith angle ($\theta>90^\circ$, in order to select only upgoing neutrino tracks. It contains 5516 neutrino candidates, with a predicted atmospheric neutrino purity of around 90\%. The median uncertainty on the reconstructed neutrino direction, assuming an  $E^{-2}$ energy spectrum, is $0.4^\circ \pm 0.1^\circ$~\cite{antares_ps2}.

\begin{figure}[!t]
\includegraphics[width=0.55\textwidth]{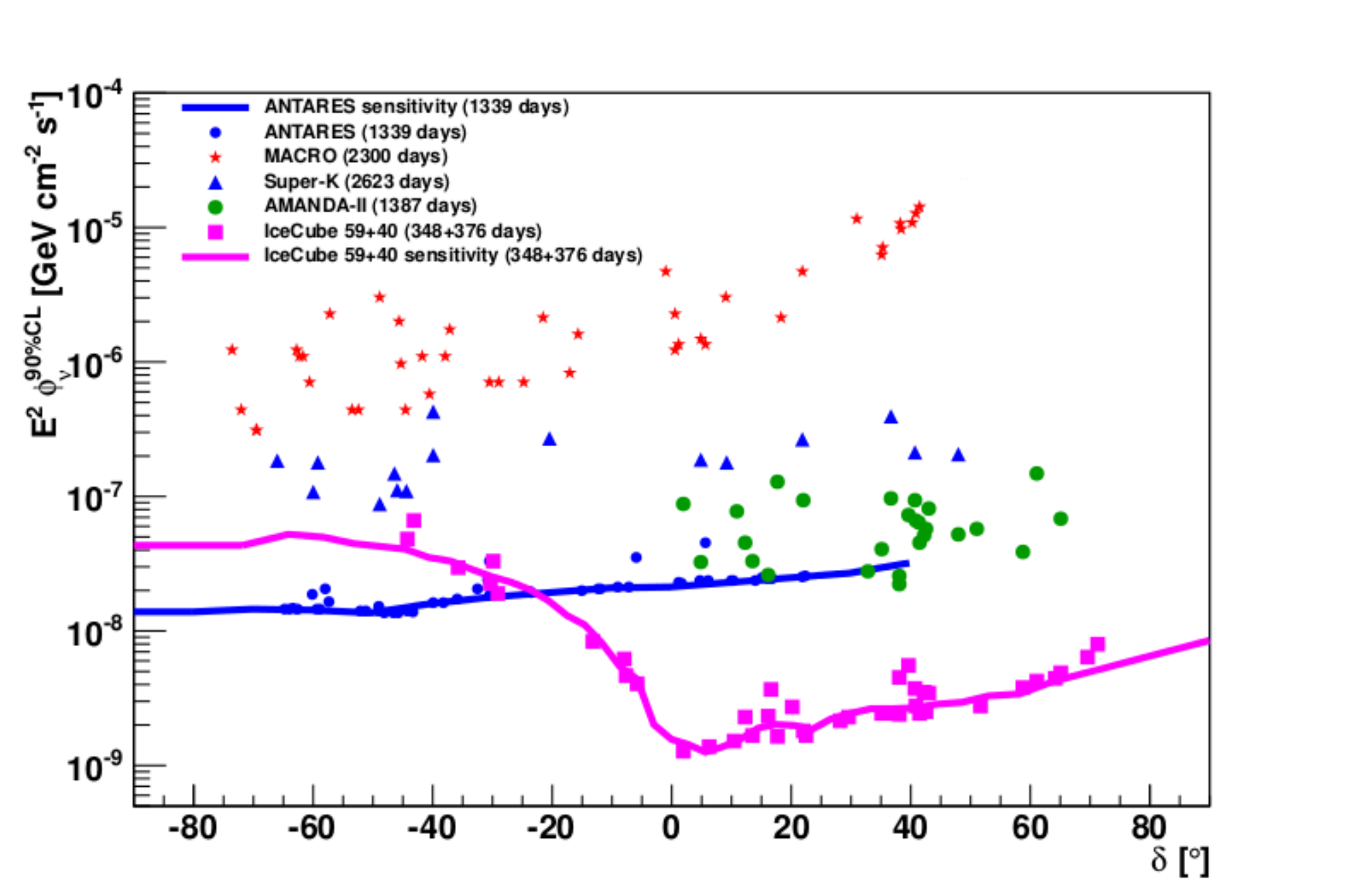}
\vspace*{-0.5cm}
\caption{ANTARES upper limits (at 90\% C.L.) on the $E^{-2}$ flux from the selected candidate sources (blue dots), and median sensitivity (blue line) as a function of the declination. Results  from other experiments are also shown, in particular the IceCube sensitivity and flux limits with combined data from the IC40 and IC59 configurations.}
\label{fig:ps}
\end{figure}

Two different searches were performed: 

-- A {\bf time-integrated full-sky search} looking for an excess of events over the atmospheric neutrino background in the declination range $[-90^\circ;+48^\circ]$. The search algorithm is based on an unbinned maximum likelihood which includes the information on the point-spread function and on the expected background rate as a function of the declination and of the number hits used in the track reconstruction. The most significant cluster of events is located at $(RA=-47.8^\circ;\delta=-64.9^\circ)$ and is the same as found in the 2007-2010 dataset; its post-trial p-value is 2.1\%, equivalent to a 2.3$\,\sigma$ significance. This "warm spot"  was also searched for counterparts in the electromagnetic band, without success~\cite{antares_mwl}.  \\
-- A {\bf candidate-list search} looking for events in the directions of a predefined list of 51 candidate sources of interest which are known gamma-ray emitters and potential sites for hadronic acceleration.   No significant excess was found in any of these 51 directions, and upper limits have consequently been derived for these candidate sources.

Fig.~\ref{fig:ps} summarizes the outcome of the above searches. The obtained limits are the most stringent  ones for a large part of the Southern Sky for sources in the TeV -- PeV energy range. IceCube limits in that region of the sky are indeed relevant only above $\sim 1$ PeV, where the downgoing muon background is effectively reduced.   

The sensitivity of the ANTARES point source search to spatially extended objects such as the shell-type supernova remnant RXJ1713.7-3946 and the pulsar wind nebula Vela X were also studied. Assuming the neutrino flux to be related to the gamma-ray flux and taking into account the measured extensions, 90\% C.L. upper limits on the neutrino flux were computed. They are respectively a factor of 8.8 (for RXJ1713.7-3946) and a factor of 9.1 (for Vela X) above the theoretical predictions (yet the most stringent ones for the emission models considered)~\cite{antares_ps2}.

\section{Multimessenger neutrino searches}
\label{sec:mm}

\begin{figure}[!t]
\hspace*{-0.8cm}
\includegraphics[width=0.52\textwidth]{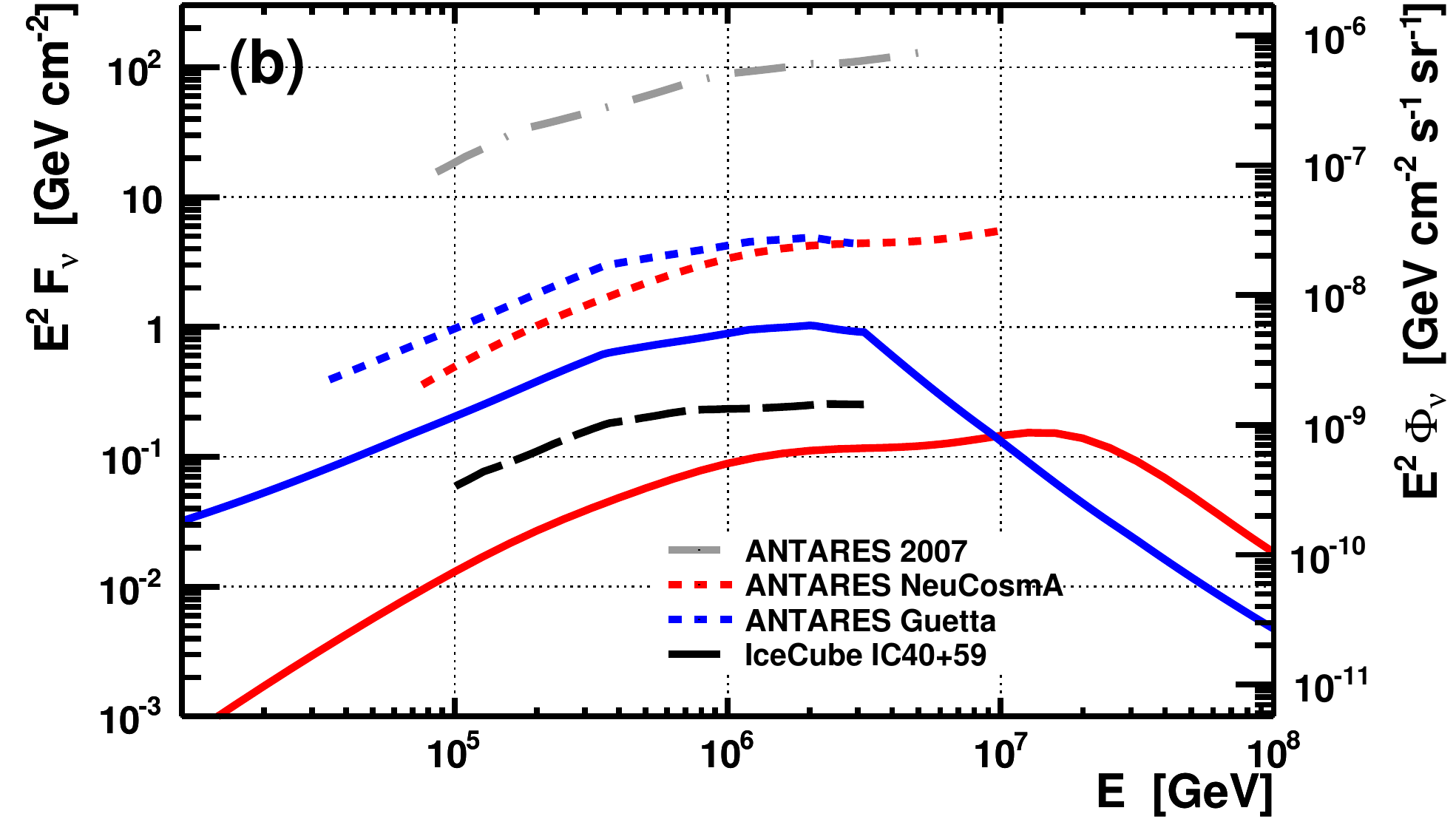}
\vspace*{-0.3cm}
\caption{Sum of the muon neutrino spectra of all selected GRBs in the 2008-2011 period (solid lines) and limits set by this analysis~\cite{antares_grb2} (short-dashed lines) as obtained from the NeuCosmA numerical simulation~\cite{neucosma} (red) or from the model from~\cite{guetta} (blue). The experimental limits~\cite{antares_grb1,ic_grb} on the total flux are also shown in grey (ANTARES) and black (IC40+59) dashed lines. The right-hand axis represents the inferred quasi-diffuse flux limit obtained assuming that each analysed sample represents
an average burst distribution and that the annual rate of
long bursts is 667 per year.
}
\label{fig:grb}
\end{figure}

Neutrino telescopes also develop specific strategies to look for neutrinos with timing and/or directional correlations with known (or potential) sources of other cosmic messengers. The limited space-time search window allows a drastic reduction of the atmospheric background, therefore enhancing the sensitivity to faint signals that would have remained undetected otherwise. This method has been used in ANTARES e.g. to put upper limits on the neutrino flux from known gamma-ray transient or flaring sources (see Secs.~\ref{subsec:grb} and~\ref{subsec:flares}).  

Alternatively, the occurrence of a special event in a neutrino telescope (such as the near-simultaneous arrival of two or more neutrinos from the same direction) could indicate that a highly energetic burst has occurred and may be used as a trigger for optical, X-ray, and gamma-ray follow-ups (see  Sec.~\ref{subsec:followup}). 

Coincident searches of neutrinos and gravitational waves are yet another example of a multi-messenger strategy that could help reveal hidden sources, so far undetected with conventional astronomy (see Sec.~\ref{subsec:gwhen}).   

\subsection{Neutrinos from GRBs}
\label{subsec:grb}

A first search for muon neutrinos in correlation with a sample of 40 GRBs that occurred in 2007, while the detector was still in a reduced, 5-line configuration, is described in~\cite{antares_grb1}. A second, more sophisticated search has been performed recently on the 2008-2011 data sample, using a set of 296 long GRBs representing a total equivalent live time of 6.6 hours~\cite{antares_grb2}. The expected neutrino fluxes are calculated for each burst individually. In contrast to the previous analyses, this search has for the first time been optimised for a fully numerical neutrino-emission model dubbed as NeuCosmA, which includes Monte Carlo simulations of the full underlying photohadronic interaction
processes~\cite{neucosma}.  The optimisation relied on an Extended Maximum Likelihood Ratio built with an a priori knowledge of the background as estimated using data off-window. 

No coincident neutrino event is found within $10^\circ$ of any of the GRBs in the sample, and 90\% C.L. upper limits are placed on the total expected flux according to the NeuCosmA model as well as to the commonly used model from~\cite{guetta}, as shown in Fig.~\ref{fig:grb}. Although not competitive with the IceCube limits~\cite{ic_grb}, the ANTARES results concern a different sample of GRBs (only 10\% of them being also in IceCube field of view), thus offering complementarity both in sky coverage and in energy range, as discussed in the previous Section.

\subsection{Neutrinos from flaring sources}
\label{subsec:flares}

To further improve the sensitivity of the detector, specific searches have been performed during periods of intense activity of some sources, as reported by X-ray and/or gamma-ray observatories:\\
-- A search for neutrinos in coincidence with 10 flaring blazars chosen on basis of their Fermi-LAT light curve profiles was performed using data from 2008. By searching for neutrinos during the high state periods of the AGN light curve, the sensitivity to these sources was improved by about a factor of two with respect to a standard time-integrated point source search. For one of these AGNs (3C279), one neutrino event was found to be in spatial ($0.56^\circ$) and time coincidence with a flare. The post-trial p-value is 10\%, and the upper limits derived on the neutrino fluence for these AGNs are presented in~\cite{antares_blazars}.  A similar analysis is ongoing on the 2008-2012 period using data from Fermi and from the Cherenkov Telescopes HESS, MAGIC and VERITAS.\\
-- A search for neutrinos in coincidence with X-ray or gamma-ray outbursts (as reported by the RXTE/ASM, Swift/BAT or Fermi/LAT satellites) of 6 microquasars was performed on 2007-2010 data. No neutrino events were detected in space-time coincidence with those flares, and the inferred limits are close to predictions~\cite{mquasars} which may be tested by ANTARES in the upcoming years, in particular for the two microquasars GX339-4 and CygX-3. \\
More detail on these analyses can be found elsewhere in these Proceedings~\cite{agustin}.

\subsection{Follow-up of neutrino events}
\label{subsec:followup}
Thanks to the ability of its data acquisition system to filter events in real-time,  ANTARES can also trigger an external detector on the basis of "golden" neutrino events selected by a fast, online reconstruction procedure~\cite{tatoo}. These events can be neutrino doublets coincident in time (within 15 minutes) and space (within $3^\circ$) or single high-energy neutrinos (typically above 5 TeV). Neutrino events closer than $0.3^\circ$ to a local galaxy (within 20 Mpc) now also generate an alert.  

This method was first proposed in \cite{followup}; the follow-up of interesting neutrino events possibly correlated to astrophysical transients can be done with fast-repositioning telescopes operating in different wavelength domains, from optical to X-rays and gamma-rays. The $2\pi$ sr instantaneous sky coverage and the high duty cycle of ANTARES are important assets for such neutrino-triggered multi-messenger programs. 

Since 2009, alerts have been sent on a regular basis ($\approx$25 per year) to a network of fast-response, wide-field ($1.9^\circ \times 1.9^\circ$) robotic optical telescopes (TAROT, ROTSE and ZADKO) and more recently also to the SWIFT/XRT telescope. The online event reconstruction allows a very short latency ($< 1$s) for the alert sending while maintaining a reasonable pointing accuracy ($\sim 1^\circ$), which is compatible with the field of view of the telescopes. Due to the repositioning time of the telescope, the first optical image can typically be obtained after a few seconds. Successive sets of images can then be recorded minutes, hours and/or days after the alert, in order to follow the time profile of different transient sources (such as GRB afterglows or core-collapse supernovae) on appropriate time scales.

No optical counterpart has been observed so far in association with one of the ANTARES neutrino alerts, and limits on the magnitude of a possible GRB afterglow have been recently set~\cite{tatoo_icrc}.

\subsection{Joint searches for neutrinos and gravitational waves}
\label{subsec:gwhen}

Both gravitational waves (GW) and high-energy neutrinos (HEN) are cosmic messengers that can traverse very dense media and travel unaffected over cosmological distances, carrying information from the inner regions of astrophysical engines such as binary mergers or core-collapse supernovae. They could also reveal new, hidden sources that have not been observed by conventional astronomy, such as the putative choked or failed GRBs. This has motivated the development of joint searches between ANTARES and the last-generation GW interferometers VIRGO~\cite{virgo} and LIGO~\cite{ligo}, which have been operating concurrently in the past few years.  A review on the motivations and challenges of joint GW+HEN astronomy can be found in~\cite{gwhenreview}. 

\begin{figure}[!t]
\includegraphics[width=0.55\textwidth]{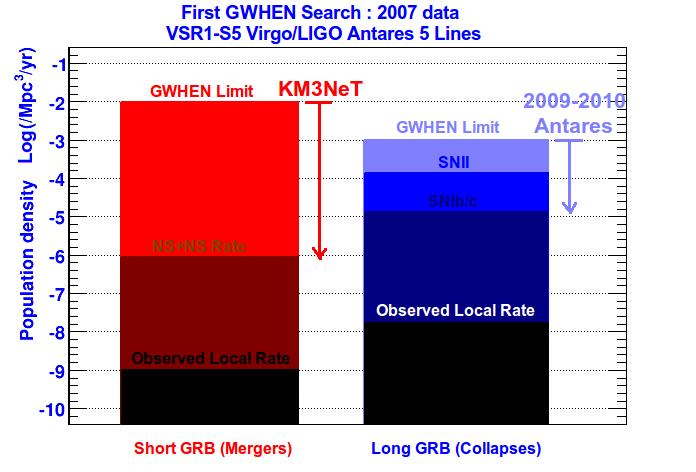}
\vspace*{-0.5cm}
\caption{GWHEN 2007 astrophysical limits on the density of two GRB populations thought to be potential GWHEN emitters, as compared to the local short/long GRB rates , to the rate of neutron star-neutron star mergers and of type II and type Ib/c supernovae. Also shown is the potential reach of future analyses with ANTARES (12 lines) and KM3NeT.}
\label{fig:gwhen}
\end{figure}

The first joint GW+HEN search uses concomitant data taken with the three detectors during 2007, during the VIRGO VSR1 and LIGO S5 science runs, while ANTARES was operating in a 5-line configuration. It consists in an event-per-event search for a GW signal correlating in space and time with a given HEN event considered as an external trigger. No coincident GW+HEN event was observed in this 2007 dataset, which allowed for the first time to place upper limits on the density of joint GW+HEN emitters~\cite{antares_gwhen}. Those limits are presented in Fig.~\ref{fig:gwhen} and compared to the densities of mergers and core-collapse events in the local universe.  

More stringent limits will soon become available from a second, optimized search using the data of the full ANTARES telescope in 2009-2010 taken in concomittance with  the VIRGO VSR2/VSR3 and LIGO S6 science runs, where all the involved interferometers took data with  improved sensitivities~\cite{icrc_gwhen}.

\section{Particle physics with ANTARES}
\label{sec:beyond}

Beyond its astrophysics reach, a neutrino detector such as ANTARES also offers broad  possibilities for particle physics. 
\subsection{Neutrino oscillations}
At the lowest end of its energy sensitivity band, ANTARES can be used for neutrino oscillation studies involving the bulk of detected atmospheric neutrinos. A measurement of the neutrino mixing parameters in the atmospheric sector has been performed with the 2007-2010 data sample, for an overall live time of 863 days.

In the simplified, two-flavor oscillation framework in vacuum, the $\nu_\mu$ survival probability can be written as
\begin{equation}
\label{eq:osc}
P(\nu_\mu \longrightarrow \nu_\mu) = 1 - \sin^2 2\theta_{23} \sin^2\frac{16200\, \Delta m^2_{23}\, \cos\theta}{E_\nu}
\end{equation}
where $\theta$ is the neutrino zenith angle (directly related to its path length across the Earth), $\Delta m^2_{23}$  is  the squared-mass difference of the eigenstates $m_2$ and $m_3$ and and $\theta_{23}$ is the corresponding mixing angle. Considering values of $\Delta m^2_{23} = 2.43 \times 10^{-3}$ eV$^2$ and $\sin^2 2\theta_{23}= 1$, as measured by the MINOS experiment~\cite{minos}, the first oscillation maximum is found for vertical upgoing neutrinos at $E_\nu = 24$ GeV, which implies a suppression of vertical upgoing muon neutrinos of such energies. 

\begin{figure}[!t]
\includegraphics[width=0.5\textwidth]{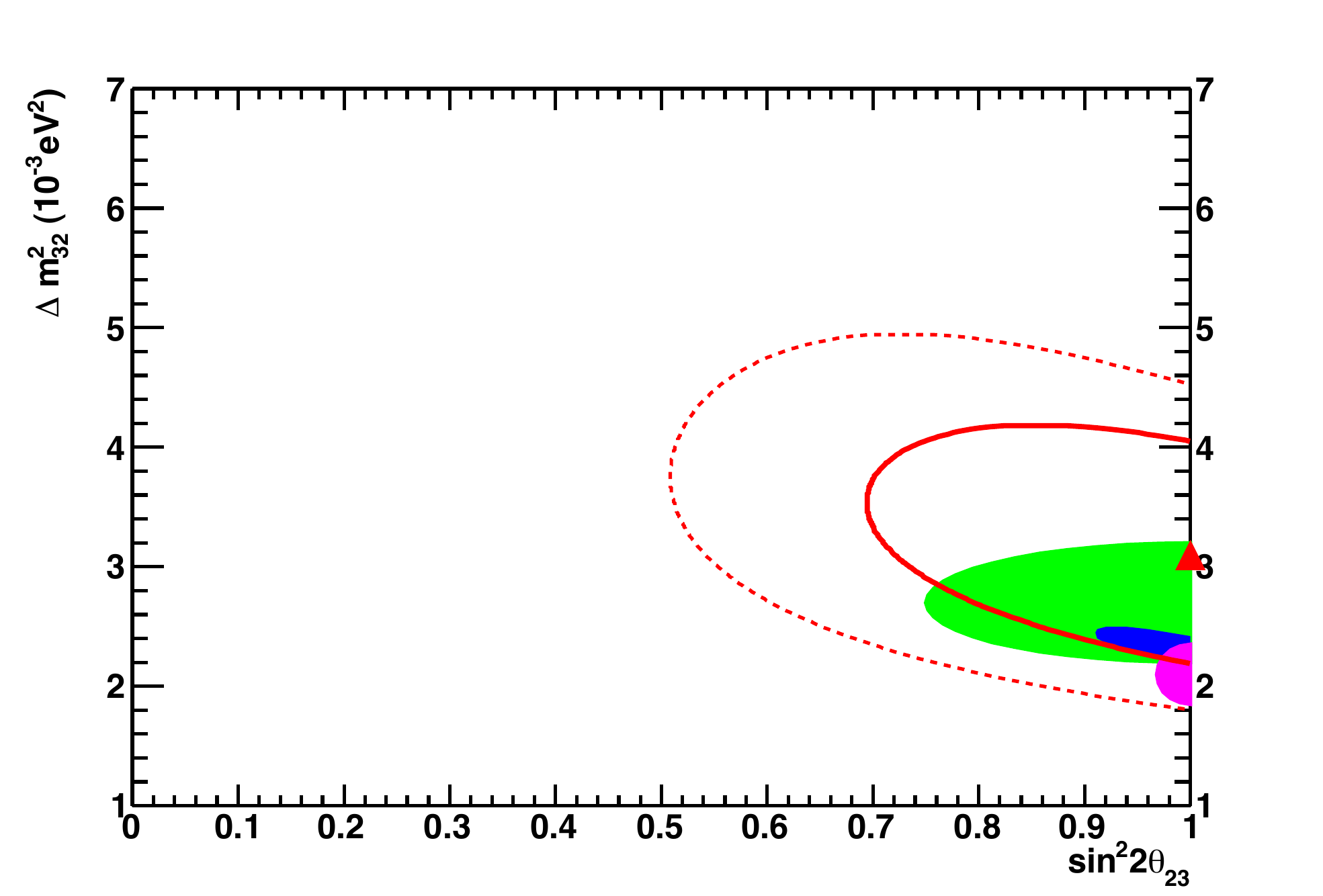}
\vspace*{-0.5cm}
\caption{68\% and 90\% C.L. contour of the neutrino oscillation parameters $\Delta m^2_{23}$ and $\sin^2 2\theta_{23}$ as derived from the fit of the reconstructed $E/\cos\theta$ distribution. The position of the best fit is indicated by the red triangle. The filled regions show 68\% C.L. contours obtained by K2K (green)~\cite{k2k}, MINOS (blue)~\cite{minos} and SuperKamiokande (magenta)~\cite{SK} for comparison.}
\label{fig:osc}
\end{figure}

The oscillation parameters are inferred by fitting the event rate as a function of the ratio of the neutrino energy to the reconstructed zenith angle, as can be seen from Eq.~\ref{eq:osc}. At such low energies, the neutrino energy can be estimated from the observed muon range in the detector. Assuming maximal mixing, a squared-mass difference of $\Delta m^2 = (3.1 \pm 0.9) \times 10^{-3}$ eV$^2$ is found~\cite{antares_osc}. The corresponding contour plot for different C.L.s is shown in Fig.~\ref{fig:osc} and compared to measurements by other dedicated experiments.

The compatibility of the ANTARES measurement with the world data gives confidence in the understanding of the detector, even close to the detection threshold. This measurement also opens the way to other studies of neutrino fundamental properties with neutrino telescopes. In particular, it has been recently pointed out that underwater/ice Cherenkov detectors could have competitive capabilities for the measurement of the neutrino mass hierarchy with atmospheric neutrinos (see e.g.~\cite{ARS,franco}). A feasibility study for such a measurment with a deep-sea detector, dubbed ORCA (Oscillation Research with Cosmics in the Abyss), is currently being performed in the framework of the KM3NeT Collaboration and briefly discussed elsewhere in these Proceedings~\cite{km3net}.   

\subsection{Dark matter searches}
ANTARES can also be used for searches of dark matter (DM) in the popularly advocated form of WIMPS (weakly interacting massive particles). Such particles get gravitationally trapped in massive bodies such as the Earth, the Sun or the Galactic Center, where they can subsequently self-annihilate, producing neutrinos with energy comparable to the WIMP mass. An excess of neutrinos in the 10 GeV -- 1 TeV energy range and in the direction of one such body could therefore provide an indirect signature of the existence of dark matter.

\begin{figure}[!t]
\includegraphics[width=0.5\textwidth]{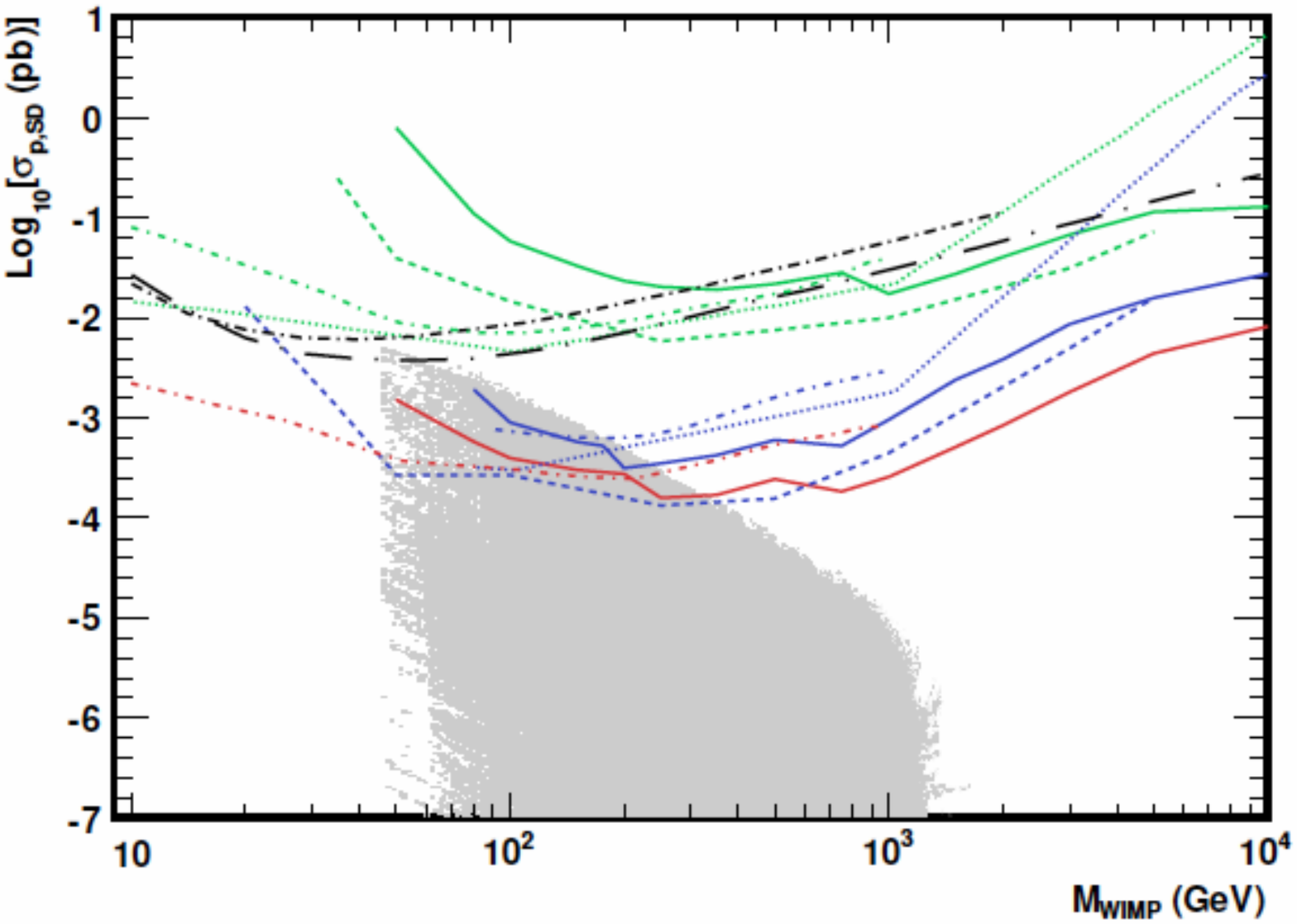}
\vspace*{-0.5cm}
\caption{Solid lines are the ANTARES 90\% C.L. upper limits on the spin-dependent WIMP-proton cross-section inside the Sun, as a function of the WIMP mass in the range $M_{WIMP}\in$ [10 GeV; 5 TeV], for the three annihilation channels: $b\bar{b}$ (green),  $W^+W^-$ (blue) and $\tau^+\tau^-$ (red). Also shown are the limits obtained from other indirect DM search experiments: SuperKamiokande 1996--2008 (dotted lines), Baksan 1978--2009 (dash-dotted lines) and IceCube 79 (dashed lines). The black lines are the limits obtained from the direct search experiments KIMS 2007 (dot-dashed) and COUPP 2011 (dashed). }
\label{fig:dm}
\end{figure}

A search in the direction of the Sun has been performed using ANTARES data recorded in 2007 -- 2008, for a total live time of 294.6 days~\cite{antares_dm}. For each WIMP mass and annihilation channel considered, the quality cuts were designed as to minimize the average 90\% C.L. upper limit on the WIMP-induced neutrino flux. No excess above the atmospheric background was found, yielding upper limits on the spin-dependent and spin-indepedent WIMP-proton cross-section; the latter ones are shown in Fig. ~\ref{fig:dm} . An improvement factor of 3 to 5 in sensitivity is anticipated in the next update of this analysis. More detail on dark matter searches with ANTARES can be found elsewhere in these Proceedings~\cite{jd_dm}.

\subsection{Searches for exotics}
Two searches for exotic particles have been performed so far:\\
-- A search for magnetic monopoles: such putative relic particles should leave a clear signature in a neutrino telescope as the intensity of light they emit is $\mathcal{O}(10^4)$ times greater than the conventional Cherenkov light, depending on their velocity $\beta$. A search for relativistic upgoing monopoles performed with 2007 - 2008 data yielded one observed event, compatible with the background hypothesis. The corresponding limits on the magnetic monopole flux for relativistic velocities $0.625 \leq \beta\leq 0.995$ are in the range $1.3 - 8.9 \times 10^{-17}$ cm$^{-2}$ s$^{-1}$ sr$^{-1}$~\cite{antares_monop}.\\
-- A search for nuclearites: these hypothetical massive particles made of lumps of u,d and s quarks could be present in the cosmic radiation either as relics of the early Universe or as debris of supernovae or strange star collisions~\cite{nucl}. They could be detected in a neutrino telescope through the blackbody radiation emitted by the expanding thermal shock wave along their path. A dedicated search, optimized for nonrelativistic, downgoing nuclearites, was performed with 2007-- 2008 data. As for monopoles, the light yield in the detector is expected to be much greater than the muon Cherenkov emission for nuclearite masses larger than a few $10^{13}$ GeV. No significant excess of such events was found in the data, yielding upper limits on the flux of downgoing nuclearites between $7.1 \times 10^{-17}$ and $6.7 \times 10^{-18}$ cm$^{-2}$ s$^{-1}$ sr$^{-1}$, for masses $10^{14} \leq M_N \leq 10^{17}$ GeV.

\newpage
\section{Conclusions}
The ANTARES detector is currently the largest neutrino telescope that has ever operated in the Northern Hemisphere, and the first to be deployed in the deep sea. Since the connection of its first detection lines in 2007, it has been continuously monitoring the Southern Sky with unprecedented sensitivity, in particular to the TeV sources lying in the central region of our Galaxy. An extended multi-messenger program complements and expands its astrophysics reach. ANTARES also paves the way to a multi-km$^3$ neutrino telescope in the Northern Hemisphere, KM3NeT, to complement the IceCube detector deployed at the South Pole. 

Beyond astrophysics,  searches for dark matter and exotic particles are also performed, and ANTARES has open the way to the study of neutrino fundamental properties with neutrino telescopes. More competitive results are expected in the future as ANTARES will continue taking data at least until the end of 2016, when it gets eventually superseded by the next-generation KM3NeT detector.

\end{document}